\documentclass[prl,twocolumn,showpacs,superscriptaddress]{revtex4}

\usepackage{graphicx,bm,amssymb,amsmath}
\usepackage{color}
\usepackage{amsmath}
\usepackage{epstopdf}

\begin{document}

\title{Athermal Phase Separation of Self-Propelled  Particles with no Alignment}

\author{Yaouen Fily}
\affiliation{Physics Department, Syracuse University, Syracuse, NY 13244, USA}
\author{M. Cristina Marchetti}
\affiliation{Physics Department, Syracuse University, Syracuse, NY 13244, USA}
\affiliation{Syracuse Biomaterials Institute, Syracuse University, Syracuse, NY
13244, USA}

\date{\today}

\begin{abstract}
We study numerically and analytically a model of self-propelled polar disks on a substrate in two dimensions. The particles interact via isotropic repulsive forces and are subject to rotational noise, but there is no aligning interaction. As a result, the system does not exhibit an ordered state. The isotropic fluid  phase separates  well below close packing and exhibits the large number fluctuations and clustering found ubiquitously in active systems. Our work shows that this behavior is a generic property of systems that are driven out of equilibrium locally, as for instance by self propulsion. 
\end{abstract}

\maketitle

Collections of self-propelled (SP) particles have been studied extensively as the simplest model for ``active materials'',  a novel class of nonequilibrium systems composed of interacting units that individually consume energy and collectively generate motion or mechanical stresses~\cite{Ramaswamy2010}.  Active systems span an enormous range of length scales, from the cell cytoskeleton~\cite{Julicher2007}, to bacterial colonies~\cite{Dombrowski2004}, tissues~\cite{Poujade2007} and animal groups~\cite{Frewen2011}.  These disparate systems exhibit common mesoscopic to large-scale phenomena, including swarming, non-equilibrium disorder-order transitions, pattern formation, anomalous fluctuations and surprising mechanical properties~\cite{Vicsek1995,Toner2005}.

Active particles are generally elongated and can order in states with either polar or apolar (nematic) orientational order.  
A remarkable property of such ordered states are giant number fluctuations. In equilibrium systems, away from continuous phase transitions, the standard deviation $\Delta N$ in the mean number of particles $N$ scales as $\sqrt{N}$ for $N\rightarrow\infty$. 
In  active systems $\Delta N$ can become very large and scale as $N^a$, with $a$ an exponent predicted to be as large as $1$ in two dimensions~\cite{Toner1995,Toner2005,Ramaswamy2003}. This  theoretical prediction has been demonstrated experimentally ~\cite{Narayan2007,Deseigne2010,Peruani2012} and verified in simulations of agent-based models~\cite{Chate2006,Chate2008,Peruani2011a}. Both nematic and polar states exhibit giant number fluctuations, which  are believed to be  associated with the  broken orientational symmetry.
% and to the interplay of density and  convective currents.
%
\begin{figure}
\centering
\includegraphics[width=0.99\columnwidth]{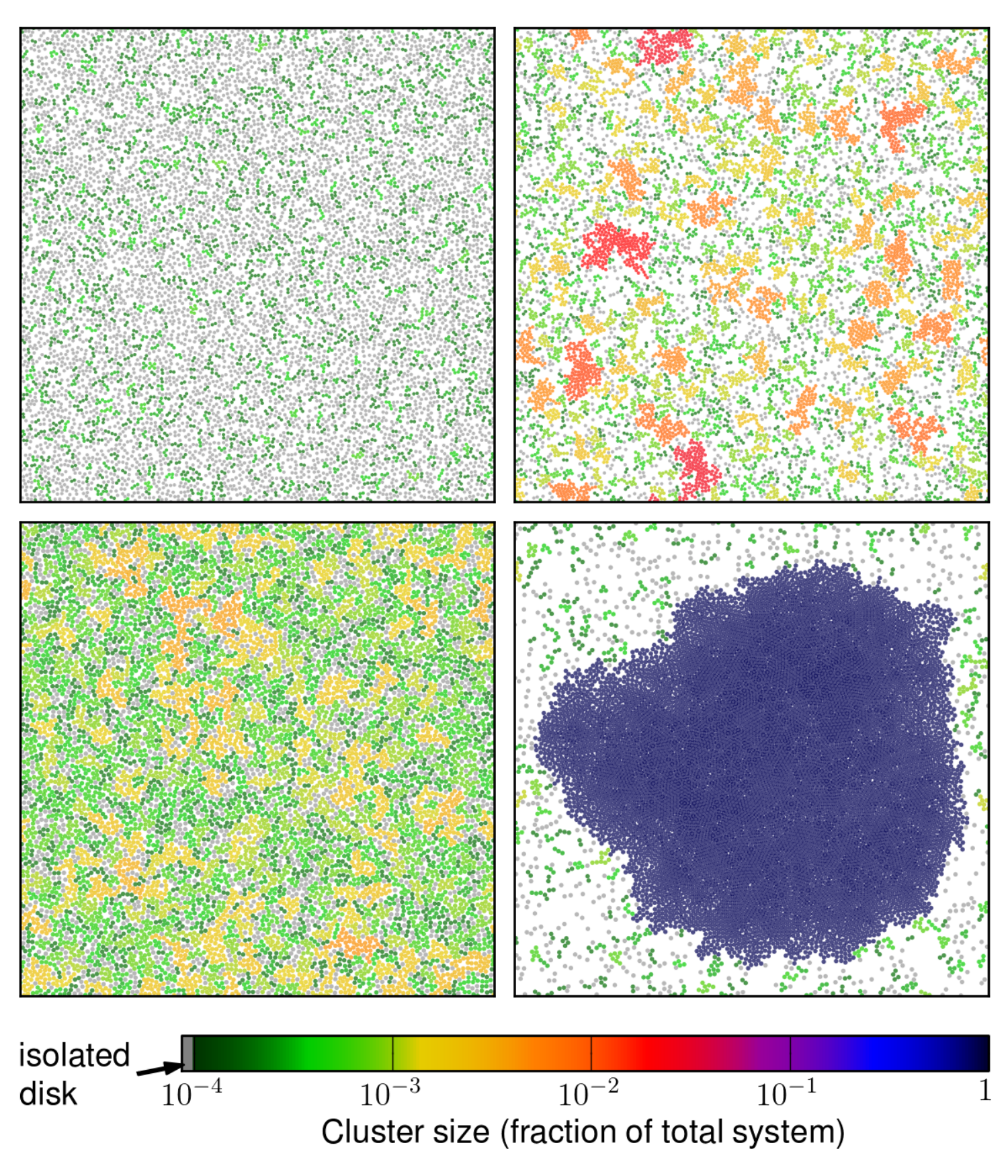}
\caption{(color online) Snapshots  of $N_T=10^4$ disks for  $\phi=0.39$ (top row) and  $\phi=0.7$ (bottom row).
Same-size clusters, defined by particles overlap, are highlighted by color coding. The left frames are for a thermal system at $k_BT=0.1$. The right frames are for SP disks with $v_0=1$ and $\nu_r=5\times 10^{-3}$. 
}
\label{fig:images}
\end{figure}
\begin{figure}
\centering
\includegraphics[width=0.9\columnwidth]{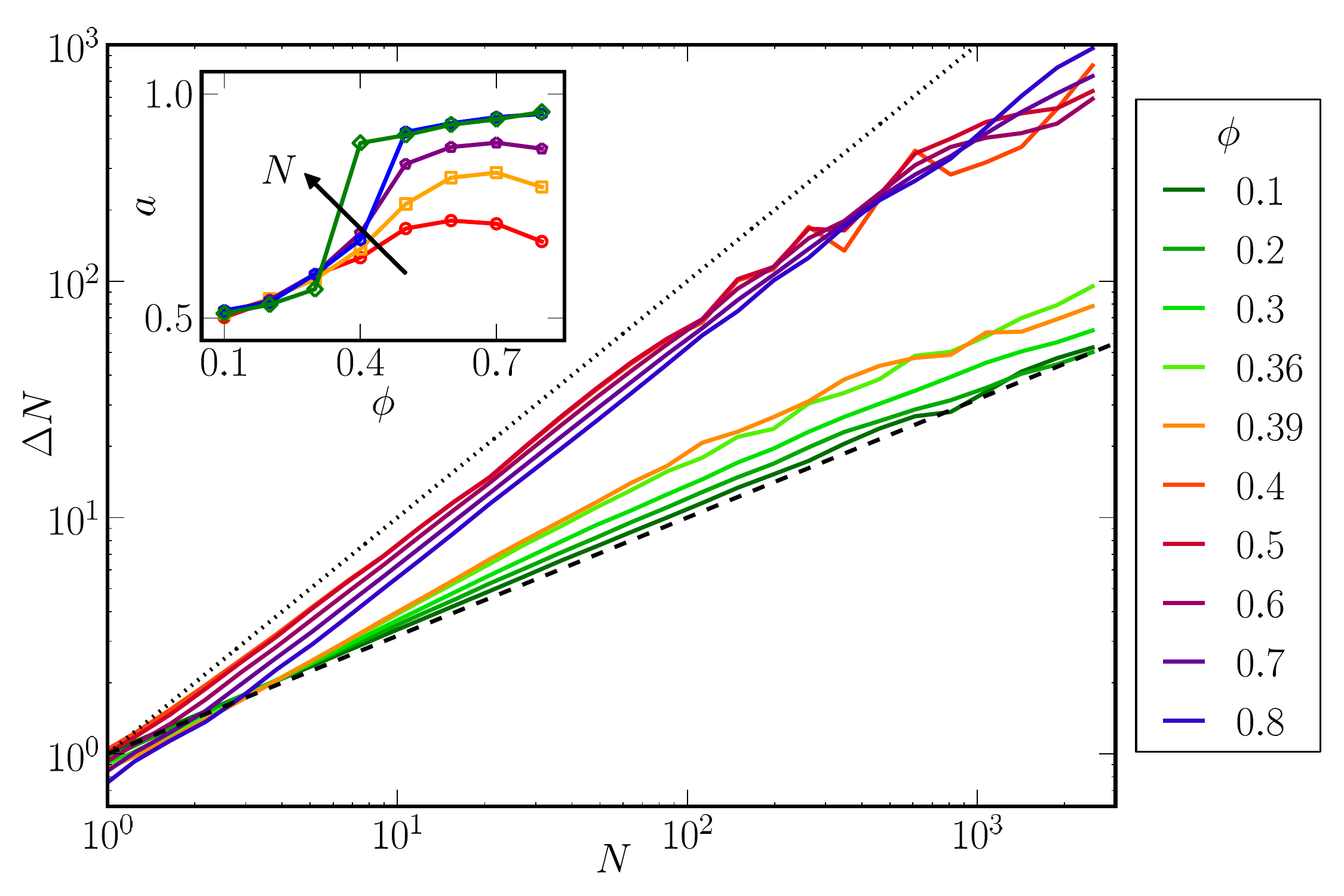}
\caption{(color online)
Standard deviation $\Delta N$ versus the average  number  $N$ of particles in a subsystem of size $\ell=\sqrt{\pi a^2 N/\phi}$
for packing fractions $\phi$ from $0.1$ to $0.8$, $N_T=10^4$ and $1\leq N\leq N_T$.
The dashed and dotted lines correspond to $\Delta N=N^{1/2}$ and $\Delta N=N$, respectively.
Inset:  exponent $a$ such that $\Delta N\sim N^{a}$ versus $\phi$ for $N_T$  from $200$ to $10^4$.
}
\label{fig:DN}
\end{figure}

In this paper we study  a model of SP soft repulsive disks with \emph{no alignment rule}. Since the particles are disks,  steric effects, although included in the model, do not yield large scale alignment, in contrast to SP rods that can order in nematic states~\cite{Baskaran2008a,Baskaran2008}. As a result, our particles, although  self-propelled, do not order in a moving state at any density. Figure~\ref{fig:images} shows, however, that above a  packing fraction $\phi_c\approx 0.4$  this minimal system phase separates  into a solid-like and a gas phase, hence  exhibits  giant number fluctuations for  $\phi>\phi_c$  (Fig.~\ref{fig:DN}). While the giant fluctuations seen in the ordered state of nematic and polar active systems~\cite{Narayan2007,Chate2006,Chate2008} are believe to be intimately related to the  broken orientational symmetry, the ones seen here arise in the absence of any broken symmetry when the rate at which  self-propulsion is suppressed due to steric trapping exceeds the rate of density convection, resulting in phase separation. Similar clustering has  been seen in spherical vibrated granular particles, although there inelasticity of the interaction may play a role~\cite{Olafsen1998,Aranson2008}.
Our work supports the suggestion by Cates and collaborators~\cite{Tailleur2008,Cates2010,Cates2012} that clustering and phase separation are generic properties  of systems that are driven out of equilibrium by a persistent \emph{local} energy input that breaks detailed balance. We further show that the notion of effective temperature cannot be used to describe the system.
% We also calculate the static structure factor $S(q)$ of the system in the region where the uniform isotropic state is stable and show that $S(q\rightarrow 0)$ diverges at a critical density well below close packing, signaling the onset of phase separation. 
Finally, using input from the numerics, we construct a continuum model that reproduces the results of the simulations.
%This  arises directly from the fact that  the system is driven out of equilibrium by an input of energy on each particle and therefore represents a generic unifying property of all active phases. 

\paragraph{Numerical Model}
We restrict ourselves to systems without momentum conservation,  such as granular materials or living organisms  on a substrate. The particles are soft repulsive disks of radius $a$ with a polarity defined by an axis $\hat{\bm\nu}_i=(\cos\theta_i,\sin\theta_i)$, where $i$ labels the particles. The
dynamics is governed by the equations
\begin{equation}
\label{model}
\partial_t{\bf r}_i=v_0\hat{\bm\nu}_i+\mu\sum_{j\neq i}{\bf F}_{ij}+\bm\eta^T_i(t)\;,~~~
\partial_t\theta_i=\eta_i(t)\;,
\end{equation}
with $v_0$  the  self-propulsion speed and $\mu$ the mobility. The translational and rotational noise terms, $\bm\eta^T_i(t)$ and  $\eta_i(t)$,  are Gaussian and white, with zero mean and correlations
$\langle\eta^T_{i\alpha}(t)\eta^T_{j\beta}(t')\rangle=2D\delta_{ij}\delta_{\alpha\beta}\delta(t-t')$ (the Greek labels denote Cartesian coordinates) and
$\langle\eta_i(t)\eta_j(t')=2\nu_r\delta_{ij}\delta(t-t')$, with $D=k_BT\mu$ the Brownian diffusion coefficient and $\nu_r$ the rotational diffusion rate. For Brownian particles of size $a$, $D\sim a^2\nu_r$ at low density.
Here, however, we treat $D$ and $\nu_r$ as independent noise strengths, with $D$ controlled by thermal noise and $\nu_r$ a measure of nonequilibrium angular noise as it may arise for instance from tumble dynamics of swimming organisms~\cite{Berg1993}. In the numerical work described below we neglect the  translational noise to  highlight the crucial role of the angular noise. The force ${\bf F}_{ij}$ between disks $i$ and $j$ is short-ranged and repulsive:
${\bf F}_{ij}=-k (2a-r_{ij}) \hat{\bf r}_{ij}$ if $r_{ij}<2a$ and
${\bf F}_{ij}= {\bf 0}$  otherwise. %with $z_i$ the numbers of neighbors of disk $i$. 
Although the self-propulsion speed $v_0$ is fixed, the instantaneous speed of each particle is determined by the forces due to the neighbors, in contrast to Vicsek-type models (but see Refs.~\cite{Szabo2006,Peruani2008,Henkes2011,Farrell2012}).   
We have performed  molecular dynamics simulations of Eqs.~\eqref{model} at $T=0$ with $N_T=100$ to
$10000$ particles and periodic boundary conditions in a box of size $L$.
We explore the phase diagram by varying the self propulsion speed $v_0$ and the
packing fraction $\phi=N_T\pi a^2 / L^2$.
In the numerics we have scaled  lengths with the
radius $a$ of the disks and times with $10/(\mu k)$, and have fixed the rotational noise strength $\nu_r=5\times 10^{-3}$.
%, such that the correlation time $\nu_r^{-1}$ of active propulsion exceeds the mean free time even at low density.  

\paragraph{Athermal phase separation.}
A hallmark property of SP particles is strong clustering in the absence of attractive interactions. This is ubiquitous in the ordered state of Vicsek-type models and has been seen even in the absence of polar aligning rules in simulations of SP hard rods~\cite{Peruani2006,Yang2010,Ginelli2010}. In this case the anisotropic shape of the particles provides an aligning, although apolar, interaction, that  enhances cluster formation. Our model, in contrast, consists of radially symmetric particles and no alignment can arise from steric effects.
Nonetheless, we observe strong athermal clustering due solely to the non equilibrium nature of the SP disks.
Above a critical packing fraction $\phi_c\approx 0.4$, the system separates into dense macroscopic clusters and a low density phase. Even below $\phi_c$  clustering is much more pronounced than in a thermal system, as illustrated on Fig.~\ref{fig:images} for packing fractions below (top) and above (bottom) $\phi_c$. The frames to the right are snapshots of the SP disks model.
The left frames show snapshots of an ``equivalent thermal system'',
defined as one  with comparable overlap between particles, at the same packing fractions.
In dimensionless units, 
the typical overlap $\delta$ in the active case is obtained by balancing the active force $v_0/\mu$ with the repulsive force $k\delta$, with  $\delta=0.1$ for $v_0=1$.
The equivalent thermal system is then obtained by setting $k_B T=k\delta^2=0.1$. This comparison indicates that the SP system cannot be simply described by an effective temperature, in agreement with~\cite{Bialke2011}, 
and in contrast to what has been argued by some authors ~\cite{Palacci2010,Loi2011}.
The inadequacy of the notion of effective temperature is also supported by an analysis of the cluster size probability distribution (not shown).
The notion of effective temperature may at best hold in very dilute systems. To see this we
formally integrate the angular dynamics and rewrite Eqs.~\eqref{model} solely in terms of translational dynamics as
\begin{equation}
\label{nonM-eq}
\partial_t{\bf r}_i=\mu\sum_{j\neq i}{\bf F}_{ij}+\bm\xi_i(t)\;,
\end{equation}
where the noise $\bm\xi_i(t)$ has zero mean and variance 
\begin{equation}
\label{xi}
\langle\xi_{i\alpha}(t)\xi_{j\beta}(t')\rangle_{\theta_0} =
2[D\delta(t-t')+\frac{v_0^2}{4}e^{-\nu_r|t-t'|}]\delta_{\alpha\beta}\delta_{ij}
\end{equation}
where $\langle ...\rangle_{\theta_0}$  includes an average over the initial values  of the angles. This shows that Eqs.~\eqref{model} are equivalent to those for  interacting soft disks with non-Markovian noise
with memory time $\nu_r^{-1}$. This noise can be approximated as white, with 
$\langle\xi_{i\alpha}(t)\xi_{i\beta}(t')\rangle_{\theta_0}=2\mu k_B\left(T+T_{e}\right)\delta(t-t')\delta_{\alpha\beta}\delta_{ij}$ and  $k_BT_{e}=\frac{v_0^2}{2\nu_r\mu}$, only for $|t-t'|>>1/\nu_r$. The effective temperature description will  be adequate only in the very dilute gas phase, when $\nu_r^{-1}$ is shorter than the mean free time. In dimensionless units this requires $\phi<\pi\nu_r^2/v_0^2$. The parameter values used in Fig.~\ref{fig:images} give $\pi\nu_r^2/v_0^2\simeq 10^{-4}$ and $k_BT_{e}=100$, i.e. essentially zero density and infinite temperature.

\paragraph{Giant Number Fluctuations and Orientational Correlations.}
In the phase separated regime for $\phi>\phi_c$ we observe giant number fluctuations: as shown in Fig.~\ref{fig:DN}
the variance $\Delta N$ of the fluctuations in the number of particles in a subregion  of size $\ell^2$, containing a total mean number of particles $N$, scales as  $\Delta N\sim N^{a}$, with $a=0.95\pm0.05$. This exponent is consistent with value $a=1$ expected for phase separation in 2d.
Orientational correlations decay exponentially in both phases and the large cluster is stationary, confirming the absence of large scale orientational order. Residual correlations exist at the surface of large clusters, where particles tend to point inward. Particles mainly enter or leave the cluster individually or in pairs. These observations demonstrate that orientational correlations, to be discussed in a future publication, do not play the central role in controlling phase separation.

%Orientational correlations are also shown in the inset to Fig.~\ref{fig:DN}. Such correlations decay exponentially in the phase separated regime confirming the absence on large scale orientational order. In addition,  the center of mass of the large cluster is stationary over the time scale of the simulations.

\begin{figure}
\centering
\includegraphics[width=0.9\columnwidth]{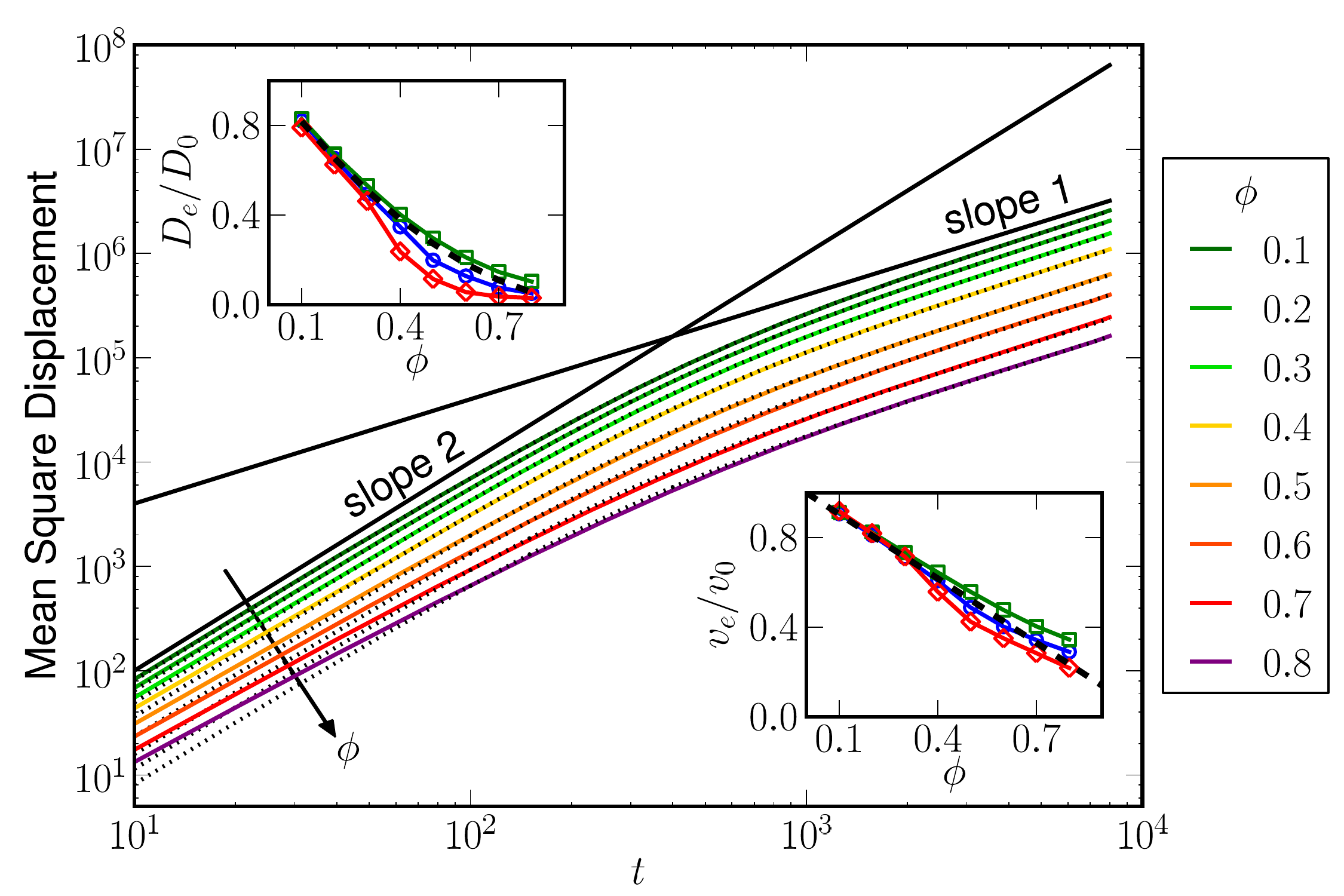}
\caption{(color online) Mean square displacement of a tagged disk versus time for various packing fractions $\phi$, showing crossover from ballistic to diffusive behavior. Insets: effective diffusion constant $D_{e}(\phi)/D_0$ and  self-propulsion speed $v_e(\phi)/v_0$ obtained from the fit to the data as functions of $\phi$ for  bare SP velocity $v_0=0.5,1,2$ (diamonds, circles and squares respectively). 
}
\label{fig:MSD}
\end{figure}

\paragraph{Mean-square displacement.}
To further characterize the dynamics, we have evaluated  the mean square displacement (MSD) of a tagged disk, shown in Fig.~\ref{fig:MSD}. An individual SP particle described by Eqs.~\eqref{nonM-eq} with ${\bf F}_{ij}={\bf 0}$ performs a persistent random walk (PRW), with 
$\langle[\Delta{\bf r}(t)]^2\rangle=4D_0\left[t+\frac{1}{\nu_r}\left(e^{-\nu_rt}-1\right)\right]$, with $D_0=\frac{v_0^2}{2\nu_r}$
and a crossover from ballistic behavior with $\langle[\Delta{\bf r}(t)]^2\rangle\sim v_0^2t^2$ for $t<<\nu_r^{-1}$ to diffusive behavior, with $\langle[\Delta{\bf r}(t)]^2\rangle\sim 4D_0t$  at long times
\footnote{Note that keeping translational noise in Eq.~\ref{model} with diffusion coefficient $D$ would yield an additional contribution $Dt$ to the mean square displacement, that we do not consider here here.}
The PRW form fits the data at vanishingly small packing fraction.
At non-zero density the MSD displacement can still be fitted by a PRW form, $\langle[\Delta{\bf r}(t)]^2\rangle=4D_{e}\left[t+\frac{1}{\nu^e_r}\left(e^{-\nu^e_rt}-1\right)\right]$, with  $D_{e}=\frac{v_e^2}{2\nu^e_r}$ and $\nu^e_r$, $v_e$ density-dependent fitting parameters. The bottom inset of Fig.~\ref{fig:MSD} shows  the effective self-propulsion speed $v_e(\phi)$  obtained from the fits as a  function of density.  
For $\phi<\phi_c$,
$v_e(\phi)$ is well fitted by a linear form $v_e(\phi)=v_0(1-\lambda\phi)$, with $\lambda \approx 0.9$ 
independent of $v_0$ for the three simulated values ($v_0=0.5,1$ and $2$).
In contrast,   $\nu_r^e\approx \nu_r$ depends  weakly on density. As a result, $D_{e}(\phi)\sim D_0(1-\lambda\phi)^2$, as shown in the top inset of Fig.~\ref{fig:MSD}.
Above $\phi_c$, the MSD is slower than ballistic at short times. In spite of this, the PRW form with a linear decrease of $v_e$ with $\phi$ still fits surprisingly well. This may be due to the fact that above $\phi_c$ a large fraction of disks belongs to stationary clusters and the MSD is controlled by particles in the low density regions. 
All densities shown here are below the crystallization density $\phi_0\approx 0.91$, above which a bounded MSD is expected~\cite{Bialke2011}.
As pointed out in Ref.~\cite{Cates2010}, $D_{e} (\phi)$ represents a collective diffusivity renormalized by interactions. This quantity was calculated in \cite{Tailleur2008} by a statistical analysis of run and tumble dynamics in one dimension, where an exponential decay of $v_e$ with density was obtained. The linear decrease of  $v_e$ with density found in our model is consistent with the predictions of Ref.~\cite{Tailleur2008} for weak interaction strength.  The resulting suppression of diffusion is  the mechanism responsible for phase separation and  cluster formation. 

\paragraph{Continuum model.}
We now use the findings from our simulations to construct an empirical continuum model that captures the dynamics of the system. The analysis of the MSD indicates that one of the effects of steric repulsion is the replacement of $v_0$ in Eqs.~\eqref{model} by $v_e(\phi)$. After this replacement, we use standard methods~\cite{Dean1996,Zwanzig2001} to coarse-grain the microscopic dynamics and derive continuum equations for 
the conserved density $\rho({\bf r},t)$ of active particles and  the polarization density ${\bf p}({\bf r},t)=\rho({\bf r},t){\bf P}({\bf r},t)$,  with ${\bf P}$  the orientational order parameter. Although our system does not order, the noisy angular dynamics of Eqs.~\eqref{model}, described in the continuum by the coupling to ${\bf P}$, is crucial in controlling the behavior of the system. A minimal version of the hydrodynamics of overdamped SP particles~\cite{Toner1995} that neglects all convective nonlinearities, but is adequate for our purpose, is given by~\footnote{A more general form of these equations  has recently been derived by coarse graining a modified Vicsek model with a density-dependent propulsion speed~\cite{Farrell2012}.}
\begin{subequations}
\begin{equation}
\label{rho-eq}
\partial_t\rho=-\bm\nabla\cdot\left(v_e {\bf p}-D_\rho\bm\nabla \rho+{\bf f}_\rho\right)\;,
\end{equation}
\begin{equation}
\label{P-eq}
\partial_t{\bf p}=-\nu_r{\bf p}-\bm\nabla (v_e\rho)
+K\nabla^2{\bf p}+{\bf f}_p\;,
\end{equation}
\end{subequations}
where ${\bf f}_\rho$ and ${\bf f}_p$ represent Gaussian white noise with zero mean and  correlations
$\langle f_{\rho i}({\bf r},t) f_{\rho j}({\bf r'},t')\rangle=2\Delta_\rho \delta_{ij}\delta({\bf r}-{\bf r}')\delta(t-t')$ and
$\langle f_{p i}({\bf r},t) f_{p j}({\bf r'},t')\rangle=2\Delta\delta_{ij}\delta({\bf r}-{\bf r}')\delta(t-t')$.
The density equation is simply a convection-diffusion equation, with advection velocity given by the local self-propulsion speed. 
We include a finite value for the density diffusion $D_\rho$ because even in the absence of translational noise in the microscopic dynamics, density diffusion (and polarization diffusion $K$) would be induced in the system through interactions. The polarization decays at rate $\nu_r$ and is convected by pressure-like gradients $\sim\bm\nabla(v_e\rho)$. 
The only homogeneous stationary state described by Eqs.~\eqref{rho-eq} and \eqref{P-eq} is the isotropic state with $\rho=\rho_0$ and ${\bf p}=0$. To examine the stability of this state  we consider the  dynamics of fluctuations $\delta\rho=\rho-\rho_0$ and $\delta{\bf p}$. Introducing  Fourier amplitudes $\left(\delta\rho_{{\bf q},\omega},\Theta_{{\bf q},\omega}\right)=\int_{\bf r}e^{-i{\bf q}\cdot{\bf r}}\int_{t}e^{-i\omega t}\left(\delta\rho({\bf r},t),\bm\nabla\cdot{\bf p}({\bf r},t)\right)$, the linearized equations take the form
\begin{subequations}
\begin{equation}
\label{rho-eq-lin}
\left[i\omega+D_\rho q^2\right]\delta\rho_{{\bf q},\omega}=-v_e\Theta_{{\bf q},\omega}
-i{\bf q}\cdot{\bf f}^\rho_{{\bf q},\omega}
\end{equation}
\begin{equation}
\label{P-eq-lin}
\left[i\omega+\nu_r+Kq^2\right]\Theta_{{\bf q},\omega}=   w q^2\delta\rho_{{\bf q},\omega}+i{\bf q}\cdot{\bf f}^p_{{\bf q},\omega}\;,
\end{equation}
\end{subequations}
where $w=v_e(\rho_0)+\rho_0v'_e(\rho_0)$. Since $v'_e(\rho_0)\equiv\left(\frac{dv_e}{d\rho}\right)_{\rho_0}<0$, $w$ can change sign, signaling self-trapping. The dispersion relations of the linear modes  are easily calculated.
At small wave vector the dynamics  is controlled by a diffusive mode
$\omega_-(q)\simeq i{\cal D}q^2$, with ${\cal D}=D_\rho+v_e w/\nu_r$ an effective diffusivity.  
If $w>0$,  convective currents associated with self-propulsion exceed self-trapping responsible for the decrease of $v_e$. Then ${\cal D}>0$ and the isotropic state is stable.  
When $w<0$ the effective diffusivity ${\cal D}$ becomes negative for $v_e|w|>\nu_rD_\rho$, signaling unstable growth of   density fluctuations and phase separation.
Using the linear fit for $v_e(\phi)$ from the numerics, we estimate that $w$ changes sign at $\phi^*=1/(2\lambda)$. If we neglect thermal diffusion ($D_\rho=0$) the isotropic state is unstable for all packing fractions $\phi>\phi^*\approx 0.45$, %comparable to 
that we identify with
$\phi_c\approx0.4$ found in the numerics. 
A finite value of $D_\rho$ shifts the instability boundary to higher density.

\paragraph{Static structure factor.}
The continuum model can also be used to evaluate the static structure factor $S({\bf q})=\frac{1}{N}\langle\delta\rho_{\bf q}(t)\delta\rho_{-{\bf q}}(-t)\rangle$, which is a direct measure of the spatial correlations of density fluctuations, with $S(q\rightarrow 0)=\frac{(\Delta N)^2}{N}$. 
We evaluate $S(q)$ in the region $w>0$  by computing the dynamical structure factor $S({\bf q},\omega)=\frac{1}{N}\langle|\delta\rho_{{\bf q},\omega}|^2\rangle$ from the linearized equations for the fluctuations with noise. Then
$S({\bf q})=\int_{-\infty}^\infty \frac{d\omega}{2\pi} S({\bf q},\omega)$.
The noise in the density equation does not contribute at small wave vectors, and we obtain
\begin{eqnarray}
\label{Sq-iso}
S(q)=\frac{v_e^2\rho_0\Delta}{[\nu_r+(D_\rho+K)q^2][\nu_r{\cal D}+D_\rho Kq^2]}\;,
\end{eqnarray}
with
$S(0)=\frac{v_e^2\rho_0\Delta}{\nu_r^2{\cal D}}$.
As the instability is approached from below ${\cal D}\rightarrow 0$ and $S(0)$ diverges, with a behavior that reminds of that at an equilibrium critical point. The decay of correlations is characterized by a length scale 
$\xi=\sqrt{KD_\rho/(\nu_r{\cal D})}\sim(\phi-\phi_c)^{-1/2}$ that diverges at the instability.  
\begin{figure}
\centering
\includegraphics[width=0.95\columnwidth]{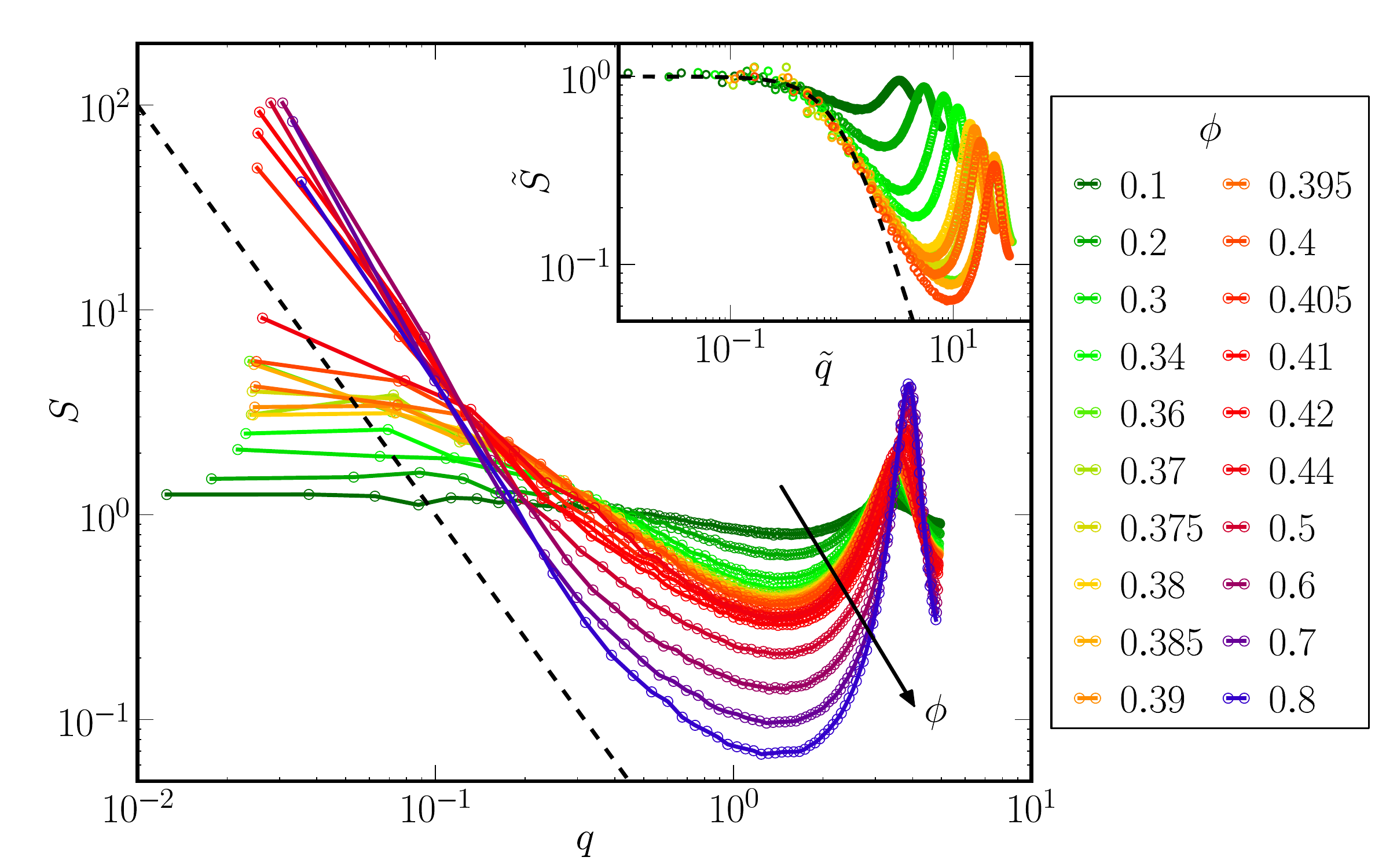}
\caption{(color online)
Static structure factor $S(q)$ for  $\phi$ from $0.1$  to $0.8$ and $N_T=2000$. The dashed curve is $S\sim q^{-2}$.
Inset: Rescaled structure factor $\tilde{S}(\tilde{q})$ for  $\phi<\phi_c$ (see text). The dashed curve is $\tilde{S}=1/(1+\tilde{q}^2)$.
}
\label{fig:SofQ}
\end{figure}
The static structure factor obtained from simulations is shown in Fig.~\ref{fig:SofQ}.
For $\phi>\phi_c$, $S(q)$ diverges at small $q$ and is reasonably well described by $S(q)\sim q^{-\alpha}$ with $\alpha\sim2$, consistent with what is expected for a phase separated system. % in the range of $q$ accessible to our simulations, 
Below the transition ($\phi<\phi_c$), we fit $S(q)$ at low $q$ to a Lorentzian $S(q)=S_0/(1+q^2\xi^2)$. %$S(q)=S_0/\left[1+(q/q_0)^2\right]$
The  inset of Fig.~\ref{fig:SofQ} shows a good collapse of the rescaled structure factor $\tilde{S}(\tilde{q})=\tilde{S}=S/S_0$, with $\tilde{q}=q\xi$ and $S_0$ and $\xi$ obtained from the fits. The growth of the correlation length $\xi$ is not inconsistent with a divergence at the transition, but 
 a detailed study of the transition region is needed to determine scaling exponents. 
 
 In summary, we have shown that self-propelled particles with no alignment exhibit an athermal clustering instability to a phase-separated regime well below close packing. Above the instability, the system exhibits large density fluctuations ubiquitous in active systems.

\vspace{0.2in}
\acknowledgments
We thank Aparna Baskaran, Silke Henkes and Davide Marenduzzo for illuminating discussions. This work was supported by the National Science Foundation through awards
DMR-0806511 and DMR-1004789.  The computations were carried out on SUGAR, a computing cluster supported by NSF-PHY-1040231, and Tapputi, a computing cluster provided by Jen Schwarz.

\end{document}